\documentclass[sigconf]{acmart}

\AtBeginDocument{%
 }

\usepackage[most]{tcolorbox}
\usepackage{xcolor}
\usepackage{tabularx}
\usepackage{booktabs}
\usepackage{array}

\copyrightyear{2026}
\acmYear{2026}
\setcopyright{cc}
\setcctype{by}

\begin{document}

\title[Stuck in a Spiral]{``Stuck in a Spiral'': Shame and Guilt as \\Social Regulators of AI Use in Computing Education}


\author{Kate Hamilton}
\affiliation{%
  \institution{Temple University}
  \city{Philadelphia}
  \state{PA}
  \country{USA}}
\email{kate.hamilton@temple.edu}
\orcid{0009-0006-7684-2871}

\author{Irene Hou} 
\affiliation{%
  \institution{University of California, San Diego}
  \city{San Diego}
  \state{CA}
  \country{USA}
}
\email{ihou@ucsd.edu}
\orcid{0009-0008-0511-7685}

\author{Dev Patel} 
\affiliation{%
  \institution{Temple University}
  \city{Philadelphia}
  \state{PA}
  \country{USA}
}
\email{devhpatel@temple.edu}
\orcid{0009-0008-6622-6703}

\author{Sheena Nnam} 
\affiliation{%
  \institution{Temple University}
  \city{Philadelphia}
  \state{PA}
  \country{USA}
}
\email{sheena.nnam@temple.edu}
\orcid{0009-0001-0777-3693}

\author{Hena Patel} 
\affiliation{%
  \institution{Temple University}
  \city{Philadelphia}
  \state{PA}
  \country{USA}
}
\email{hena.patel0002@temple.edu}
\orcid{0009-0009-0570-6014}

\author{Stephen MacNeil} 
\affiliation{%
  \institution{Temple University}
  \city{Philadelphia}
  \state{PA}
  \country{USA}
}
\email{stephen.macneil@temple.edu}
\orcid{0000-0003-2781-6619}

\renewcommand{\shortauthors}{Hamilton et al.}

\begin{abstract}

While prior work has examined patterns of adoption and social norms around AI use, less is known about how emotional factors, such as shame and guilt, shape students use of AI tools. 
We present an interview study with 19 computing students through a functionalist perspective of shame and guilt, which interprets emotions as social signals that regulate behavior. Our findings show that these emotions regulate when and how students make their use visible, as they engage in hiding behaviors and selective disclosure. Students described shaming themselves, their peers, and even faculty for using AI. Shame and guilt often coexist with continued AI use, creating cycles of reduced agency and moral tension rather than promoting behavior change. Students described feeling tensions between their AI use and their identities as competent, hardworking, or ethical computing students. Students also used language and metaphors of addiction to describe their experiences. These results highlight the need to consider the socio-emotional aspects of AI use, which may be influenced by how AI policies are implemented and enforced. We discuss classroom practices that can foster healthy, open discussion and support responsible AI use.

\end{abstract}




\maketitle



\section{Introduction}
Students are widely adopting generative AI (genAI) in the classroom~\cite{hou2024effects, hou2025usage, alpizarChacon2025excited} due to its ability to provide high-quality support~\cite{prather2023robots, prather2025beyond} that can be tailored to their specific needs~\cite{bernstein2024like, logacheva2024evaluating}. Many courses are adapting to incorporate AI in classwork and assignments~\cite{vadaparty2024cs1, reeves2025prompts}. At the same time, recent work by Hou et al. has highlighted how the growing use of AI for personal help may be displacing support from peers and interfering with learning networks~\cite{hou2025all}. Preliminary evidence shows that students experience guilt about their use of AI tools, with implications around how these emotions may harm how students interact and engage within the classroom~\cite{hou2025all}. However, despite rising concerns, a recent review of the harms and unintended consequences of genAI in computing education indicate that social harms are least investigated compared to the cognitive, metacognitive, and logistical harms of AI~\cite{bernstein2025beyond}. 

Understanding social harms requires looking beyond what students do with AI to how they feel about using it and how they believe others perceive that use. Thus, in this paper, we explore the socio-emotional experiences that shape how students engage with AI in computing courses. Rather than focusing solely on patterns of usage, which have received extensive attention~\cite{hou2024effects, hou2025usage, alpizarChacon2025excited}, our work examines students' subjective experiences and how they interpret, justify, and make sense of their AI use in relation to peers and themselves. 
We conducted semi-structured interviews with 19 computing students across four R1 universities about their experiences using AI in their coursework. Our analysis was guided through a functionalist perspective of guilt and shame~\cite{tangney2003shame, hooge2011function}, which conceptualizes these emotions as social signals that regulate behavior, relationships, and identity. From this perspective, feelings of shame and guilt do not simply reflect students' discomfort, but also play an active role in shaping how students disclose, conceal, or rationalize their use of AI. Guided by this framing, we asked the following research questions: 

\begin{itemize}
    \item [\textbf{RQ1:}] How do students describe their experiences using generative AI in computing classrooms?
    \item [\textbf{RQ2:}] How do feelings associated with guilt and shame emerge in students' narratives about generative AI use? 
    \item [\textbf{RQ3:}] How do these emotional experiences shape students' behaviors, identities, and social interactions around learning? 
\end{itemize}

Our findings provides preliminary evidence that students' experiences with genAI are deeply entangled with social judgment and their identity. Students described hiding or downplaying their AI use, engaging with tools secretively, and avoiding discussions about AI with peers or instructors. Many participants described internalizing negative self-assessments about their AI use by characterizing themselves as lazy, dishonest, or less capable. Others described instances of peer judgment and public shaming. In some cases, students described their use as addictive and accompanied by cycles of secrecy, shame, and guilt that discouraged open discussion and transparency. 

Our work expands the recent growing focus on the social aspects of AI use~\cite{hou2024effects, hou2025all, padiyath2024insights, alpizarChacon2025excited}. Our findings suggest that social concerns are not limited to social displacement~\cite{hou2025all} or social shaping~\cite{hou2025all, padiyath2024insights}. AI is also a social mediator peer perceptions and computing identities.

\section{Functionalist Perspectives of Shame and Guilt}


Prior work has investigated whether and how students adopt AI tools including norms around their use in class. Students increasingly adopt these tools~\cite{alpizarChacon2025excited, hou2024effects, hou2025usage}, sometimes displacing peer interactions and negatively impacting learning communities~\cite{hou2025all}. Cultural factors also seem to impact students' adoption~\cite{alpizarChacon2025excited} with younger students being more willing to use AI. Social shaping theory suggests that students' use of AI is strongly influenced by their perceptions of how others use these tools~\cite{padiyath2024insights}. While these studies highlight social influences, they provide less insight into the emotional mechanisms that regulate AI use in classroom contexts. 


To interpret these dynamics, we draw on a \textbf{functionalist perspective of  social emotions}, which conceptualizes emotions as adaptive regulators of behavior in social contexts~\cite{tangney2003shame, hooge2011function}. In this perspective, guilt is often oriented toward a specific behavior (``I did something wrong'') and motivates reparative or corrective action~\cite{tagney1996shame, lewis1975shame}, whereas shame is oriented toward the self (``I am wrong'') and elicits withdrawal, concealment, or threats to identity~\cite{lewis2008self, smith2002public}. Both emotions can co-occur but have distinct behavioral consequences.

In educational contexts, guilt and shame can influence help-seeking, self-perception, and self-regulation. Prior research suggests that experiences related to shame and guilt, which are often expressed through fear of negative evaluation, self-doubt, and identity threat, are all associated with reduced participation and sense of belonging~\cite{maymon2018when, huff2016exploring}. These associations are concerning because identity and sense of belonging have been consistently correlated with student persistence, retention, and performance~\cite{allen2008third, salguero2021understanding, garcia2023regular, tonso2006student, mahadeo2020developing, falkner2015gender}.

Prior work in computing education has studied the emotions (e.g. frustration, confusion, boredom) that students experience while programming and how they relate to learning and performance~\cite{lishinski2022self, rodrigo2009affective, bosch2013emotions}. Feelings of frustration, annoyance, and boredom can negatively impact student learning outcomes and performance~\cite{coto2022emotions}, as well as self-efficacy and identity formation in computing\citet{kinnunen2012my}. However, less is known about social emotions, or the relational dimensions of students' experiences that emerge from their concerns about how they are perceived, judged, or evaluated by others. Unlike emotions like frustration or boredom, shame and guilt are self-conscious emotions that derive from perceived violations of personal or social standards~\cite{tangney2003shame}.

In the context of AI use, these emotions become very consequential. Students must continually interpret whether their use of AI is responsible, legitimate, or potentially harmful to themselves or others, while also anticipating how their use will be interpreted by peers and instructors. For instance, students may feel guilty if they think their use of AI gave them an unfair advantage over their peers (e.g. ``cheating,'' unauthorized use). Students may feel shame if their AI use conflicts with their sense of themselves as competent learners. As students and instructors together negotiate norms around responsible AI use in computing education, it is critical to understand shame and guilt in the context of students' AI use. This understanding can inform the design of AI policies, pedagogical interventions, and classroom strategies.

\begin{table*}[h!]
\newcolumntype{R}{>{\raggedleft\arraybackslash}X}
\centering
\caption{We interviewed 19 undergraduate computing students from four R1 universities across the United States. The `Year' column indicates current university standing. The `Frequency' column represents reported frequency of AI use.}
\label{tab:participant-demographics}
\setlength{\tabcolsep}{10pt}
\renewcommand{\arraystretch}{1.25}
\small
\begin{tabularx}{0.77\textwidth}{lllXl}
\toprule
\textbf{ID} & \textbf{Gender} & \textbf{Year} & \textbf{Major (Minor)} & \textbf{Frequency} \\
\midrule
P1  & M  & 2nd-year & Computer Science & Hourly \\
P2  & W  & 2nd-year & Information Science and Technology & Daily/Weekly \\
P3  & W  & 4th-year & Data Science & Daily \\
P4  & M  & 4th-year & Computer Science & Daily/Weekly \\
P5  & M  & 2nd-year & Computer Science/Math Economics & Daily \\
P6  & W  & 3rd-year & Computer Science & Bi-Weekly \\
P7  & W  & 4th-year & Computer Science (Math) & Daily \\
P8  & W  & 3rd-year & Computer Science & Bi-Weekly \\
P9  & W  & 2nd-year & Computer Science & Bi-Weekly \\
P10 & W  & 4th-year & Economics (Computer Science/Psychology) & Daily \\
P11 & M  & 3rd-year & Computer Science & Daily \\
P12 & M  & 2nd-year & Computer Science & Daily \\
P13 & M  & 1st-year & Computer Science & Daily \\
P14 & M  & 1st-year & Computer Science & Daily \\
P15 & NB & 1st-year & Computer Science & Daily \\
P16 & W  & 4th-year & Psychology (Information Science and Technology) & Weekly \\
P17 & W  & 2nd-year & Information Science and Technology & Weekly \\
P18 & W  & 3rd-year & Computer Science/Psychology (Criminal Justice) & Daily \\
P19 & W  & 3rd-year & Computer Science & Weekly \\
\bottomrule
\end{tabularx}
\end{table*}

\section{Methods}

To understand computing students' experiences with AI tools, we conducted 19 semi-structured interviews guided by a functionalist perspective~\cite{tangney2003shame, hooge2011function} of social emotions such as shame and guilt. Semi-structured interviews aimed to elicit participants' interpretations of emotionally meaningful experiences and how these emotions influenced their behaviors and social interactions. While our interview questions focused broadly on students' AI use, motivations, and strategies, this framework informed our interpretive lens, allowing us to identify relevant students' behaviors, such as selective disclosure. Interviews are commonly used to study emotion in computing education. For example, \citet{coto2022emotions} found that nearly half of computing education studies on emotion relied on self-report methods, including interviews.
Consistent with inductive qualitative research, our goal was to surface and interpret students' lived experiences rather than test predefined hypotheses.

\subsection{Interviews}

We conducted 30-45 minute recorded interviews via Zoom with verbal and written consent. Participants were discouraged from sharing their screens or videos to protect their privacy. Participants were compensated with a \$10 gift card, and this research was approved by our university's Institutional Review Board (IRB). 

\subsubsection{Participants and Recruitment}

We recruited 19 undergraduate students from four R1 universities across the United States. Students were recruited via announcements by computer science faculty and computing-related student organizations. Interviews were conducted in 2025 between July and November. The demographic information for participants is reported in Table~\ref{tab:participant-demographics}. Participants were diverse in gender (7 men, 11 women, 1 non-binary), university level (3 first-year, 6 second year, 5 third year, 5 fourth year), and computing sub-area. Their AI usage ranged from hourly (1) and daily (10), to weekly (8). Participants also described experiencing diverse AI policies ranging from highly permissive to highly restrictive. Policies often varied widely from course to course even for the same participant. 

\subsubsection{Interview Protocol and Rationale}

To understand students' emotional experiences, our interview questions were informed by functionalist perspectives of shame and guilt~\cite{tangney2003shame, dempsey2017function, barrett1995function, hooge2011function}. Interviews were chosen as a method because shame and guilt are often internal processes which can lead to concealment and avoidance, and therefore are not often observable through behavior alone. Interviews have also been used frequently to study emotions in computing education contexts~\cite{coto2022emotions}. Our focus is on understanding how students interpret and narrate their behaviors and emotions in the context of AI use, but not on measuring the presence or intensity of emotions. 

Accordingly, we avoided directly asking about specific emotions, and designed questions that focused on students' perceived responsibilities (to themselves and to others), social norms, and role expectations. This allowed emotional experiences to emerge through students' own accounts of behaviors and interpretation. 

Interviews were semi-structured to build rapport given the sensitive and potentially stigmatizing nature of the discussed topics. Participants were initially asked about their year in school, major, and how often they use AI tools. Next, we asked students to compare how comfortable they felt disclosing their AI use to people in different roles (e.g. peers, professors, TA's) to surface role-based differences in disclosure. Subsequent questions asked students to reflect on moments when someone saw them using an AI tool. This naturally elicited participants to reflect on their emotional reactions, the reasoning behind their choices, and their strategies for managing visibility or judgment. Finally, students were asked about their perceived responsibilities as university students in using AI and how these responsibilities influenced their feelings about AI use.
Questions were framed around behaviors in a nonjudgmental way to avoid biasing participants' responses and allow participants to share experiences at their own comfort level.

\subsubsection{Peer Interviews}

To mitigate power asymmetries and encourage candid reflection, the interviews were conducted by trained undergraduate researchers who were peers of the participants. This peer interview approach was intentional because we anticipated that discussions of AI use, shame, and guilt might be constrained in interviews with faculty or graduate students. Undergraduate students received extensive training in qualitative interviewing, ethical research practices, and reflexivity.

\subsubsection{Interview Analysis}

The interview data was analyzed using a theoretically informed, primarily inductive approach. The analysis was guided by a functionalist framework of shame and guilt~\cite{tangney2003shame, hooge2011function} and informed by prior work on these constructs~\cite{pivetti2015shame, teroni2008shame, baldwin2006relationship}. The framework guided our focus on how students evaluated their behaviors and identities in relation to their AI use. Within this framing, we adopted an inductive approach to remain open to unanticipated patterns in students' experiences, rather than imposing predefined categories of shame and guilt. The framework helped guide, but did not constrain, the organization of the codes. 

Four researchers independently applied open codes to the interview transcripts, and they met regularly to discuss excerpts from the transcripts, compare codes, and develop a shared understanding of the data. These discussions were not aimed at identifying a single ``correct'' coding, but to deepen the team's collective understanding of students' experiences through comparisons of perspective. 

Themes were developed iteratively through multiple rounds of coding and discussion. The early rounds focused on generating and comparing codes, and the later rounds involved grouping related codes, identifying broader patterns, and refining boundaries for the themes. Themes were constructed based on their explanatory power across participants and their coherence as a theme. The team paid special attention to instances that challenged or did not fit the pattern of the theme. These instances were used to refine the themes and contextualize observed patterns.  

Reflexivity~\cite{Clarke04052017} was an important component of the analysis. Individuals reflected on how their perceptions and use of AI shaped their interpretations. To address power dynamics in the team, senior researchers prioritizing facilitation and deferred initial interpretations to more junior members to create space for diverse perspectives. The team also repeatedly revisited earlier interpretations in light of new insights and interpretations of the data.

\subsection{Positionality Statement}

The research team maintained an open stance toward AI tools, recognizing both the benefits and the concerns raised in prior work. While this work was motivated by emerging evidence of shame and guilt-related experiences around AI use~\cite{hou2025all}, we did not presuppose students would experience these emotions. In interviews, we aimed to elicit students' personal experiences, feelings, and behaviors in their own terms. Follow-up questions remained open-ended and focused on clarification and sensemaking rather than evaluation. We explicitly avoided imposing normative judgments about AI use or framing practices as inherently appropriate or inappropriate.

\section{Results}

\subsection{Hiding Behaviors and Selective Disclosure}


When asked about when they felt comfortable sharing their use of AI, a subset of participants (8 of 19) described selective disclosure that was limited to specific social contexts. Consistent with prior work on social shaping~\cite{padiyath2024insights}, this subset of participants reported feeling comfortable sharing their AI use with close friends or in situations where others were already using AI tools openly. However, many of them also emphasized that such disclosure did not extend beyond these narrow contexts.

Across the broader sample, and including participants who described limited disclosure with trusted peers, students overwhelmingly reported concealing their AI use from classmates and instructors. This suggests that disclosure and and concealment are not mutually exclusive behaviors.

\subsubsection {Hiding Behaviors}

Students often hid their use, including closing the tabs, dimming the screen, etc. For example, P6 said,  

\begin{quote}
    \textit{``I try to ask [ChatGPT,] and then I quickly get off of it and \textbf{close the tab}.''} (P6)
\end{quote}

\noindent Other students selectively hid their use based on who was present, such as in the presence of authority figures. For example, P12 switched tabs because they felt embarrassed using AI near their professor and changed the tab to hide their use:  
\begin{quote}
    \textit{``If [my professor is] walking behind me [and] I have [ChatGPT] up, trying to summarize or explain the text to me, I normally \textbf{shift the tab over}, so \textbf{I do feel I guess a bit embarrassed when I use it.}''} (P12)
\end{quote}

\noindent Another hiding behavior was lowering the brightness of the screen around others. For example, P10 said, \textit{``I would definitely be \textbf{turning my screen brightness way down}.''}

\subsubsection{Disclosure}

These hiding behaviors coincided with a general reluctance to disclose their use unless they observed others using it. Students described feeling guilty about these behaviors, but it did not push them to disclose their use. For example, P17 said,

\begin{quote}
    \textit{``Especially like if you're in a conversation with [a peer] and they're talking about how they didn't do good, but then \textbf{you know yourself that you used AI. So you kind of feel guilty... and awkward}.''} (P17)
\end{quote}

\noindent Hiding behaviors and disclosing behaviors did not appear to be mutually exclusive; they were each driven by the context of who was around a student and their perceived social standing, whether it was peers or perceived `others.' This aligns with prior findings on selective disclosure of AI use~\cite{hou2025all}.

\subsection{Shaming Oneself and Others}


Beyond selective disclosure, participants described how AI use was often entangled with shame. Shaming was frequently directed at oneself, peers, and in one case, a faculty member. This culture of shaming appeared to shape how students evaluated themselves, monitored others' behavior, and regulated when and where AI use felt acceptable.


\subsubsection{Shaming Oneself}

Many students reported feeling frustrated with themselves after using AI, even when that use aligned with course policies. For example, P14 voiced a persistent pattern of negative self-talk about their AI use, 

\begin{quote}
    \textit{``\textbf{It always feels bad}. Even if I'm trying to only get an example problem, I feel it's just doing everything for me. \textbf{I'm always talking down on myself...like, I wish I was doing this myself}.''} (P14)
\end{quote}

\noindent Similarly, P12 framed their emotional response to AI use in terms of regret and reduced confidence: 

\begin{quote}
    \textit{``I guess \textbf{disappointed a little bit in myself and also regretful}... it was kind of showing that I didn't have confidence in myself or my own abilities.''} (P12)

\end{quote}

\noindent Across these accounts, shame amplified students' awareness of their AI use and its perceived moral implications. However, shame did not appear to function in a way that improved the self-regulation of their behaviors. 

\subsubsection{Shaming Others} Shame also emerged through social interactions with others where students described either feeling judged or judging others for their AI use. 

\begin{quote}
    \textit{``The girl next to me saw that I was using AI and she was like, `What are you doing are you using AI?' I felt called out, \textbf{like I was caught red-handed... I felt the need to hide it away immediately}.''} (P15)
\end{quote}

\noindent While many accounts described being personally shamed or observing others being shamed, a few participants also spoke about shaming others or judging them for their AI use. For example,  

 \begin{quote}
     \textit{``For me, \textbf{when I see people use ChatGPT in my classes, I lowkey judge them} from the back, so in front of strangers, I would not want to use it.''} (P6)
 \end{quote}

\noindent While shame was often directed toward peers, there was one instance where shame was directed toward an authority figure as well. P15 described the scenario thusly, 

\begin{quote}
    \textit{``\textbf{Our teacher did use AI} for references to come up with ideas, and \textbf{all the students would shame him} ... so it was kind of \textbf{nerve-wracking to talk about it around these people because then they're gonna shame you} and they're gonna jump on you.''} (P15)

\end{quote}

\noindent These dynamics sometimes extended beyond classrooms into students' personal lives. In one such example, P16 said she felt uncomfortable when her sister saw her using AI, 

\begin{quote} 
    \textit{``I used AI for help, and then \textbf{my sister saw and was like, `Oh no, not that.'} I guess I felt, like, shame about it. I was like, `Oh'.'' } (P16)
\end{quote}

\noindent P16's reaction did not stem from an explicit accusation or rule violation. Light expressions of disapproval from close others, like a family member, also appeared to  evoke shame.

\noindent Students described how public and private judgments can discourage use. Judgments came from peers, friends, and even family; however, it is unclear how the strength of these social ties impacts the feelings of shame that students described. The shaming of a perceived authority figure suggests that such judgments are at least partially normalized. As a result, judgment operates as both an experienced and enacted social process. These public judgments may make it more difficult for students to openly discuss AI use.

\subsection{Shaping Identity}

Students in interviews talked about how their AI use and perceptions of their use impacted their computing identities. For example, P10 described themselves as a good student and provided evidence by referencing their
honors program. They went on to describe being perceived as one `\textit{those} students' who uses AI as a threat to their identity. 

\begin{quote}
    \textit{``I'm a good student. I'm part of the honors program. I don't want my teacher to think I'm one of \textbf{those} students.''} (P10)
\end{quote}

\noindent In another case, the student explicitly described withdrawing from the class and not wanting to be perceived by the professor due to their shame in using AI, 

\begin{quote}
    \textit{``I could get tutoring [at] my professor's office hours, but \textbf{I feel kind of hesitant to go there because it's more personal and he's seen me in class.} I feel like he knows that I'm struggling, but it's harder to talk to him in person. And so I've been using AI to help me.''} (P17)
\end{quote}

\noindent Across the interviews, students often described threats to their identity. They described not wanting to see themselves or been seen as dishonest, lazy, or incompetent. They also described a tension between their use and these negative labels.

\subsubsection{Identifying as a cheater}

Students described feeling like `cheaters' for using AI, even in courses where AI was permitted. For example, P14 explained that despite none of his teachers explicitly banning AI, he still associated AI use with cheating: \textit{``When I see someone use AI, the first thing I think of is like, cheating.''}

P9, similarly, despite professors of hers allowing AI use, sees it as "cheating yourself" when relying too much on AI usage: 
\begin{quote}
    \textit{``I think honestly using AI, even though like our professors even explicitly do allow it at times, there's still a \textbf{connotation that using AI is kind of cheating}... if you're using it for the logic and thinking part, it's a little bit like you're cheating yourself.''} (P9)
\end{quote}

\subsubsection{Identifying as lazy}
P15 felt frustrated with themselves after using AI. They also expressed guilt that they did not complete the work independently. 
\begin{quote}
   \textit{``I found myself using AI, and I felt really guilty that I should have known how to do this, \textbf{or at least put in more time looking into how to do it.}'' } (P15)
\end{quote}

\subsubsection{Identifying as incompetent}

A few participants indicated that using AI threatened their identity of being an `actual programmer.' For example, P15 said, 

\begin{quote}
    \textit{``\textbf{I do really like programming and coding, [but] when I would start to use AI, I felt like a fraud}, like I wasn't actually a programmer.''}
\end{quote} 

\noindent P8 more explicitly described this feeling as imposter syndrome: 
\begin{quote}
    \textit{``... in front of random classmates, I tend to avoid it. I guess it's \textbf{low-key a bit of imposter syndrome}.'' } (P8)
\end{quote}

\noindent These examples provide further evidence to extend prior work where students described themselves as `stupid', `foolish', and `lazy' for relying on AI tools~\cite{hou2025all}. These findings suggest that students may internalize AI use as evidence of incompetence, leading them to question their legitimacy as programmers and contributing to feelings of imposter syndrome. Some researchers have even described accusations of AI use as a classist slur~\cite{sarkar2025classist}.

\subsection{Cycles of Use}

Across interviews, participants frequently described recurring patterns of AI use that persisted despite feelings of guilt or concern about its impact on their learning. Rather than acting as a deterrent, these negative emotions often became part of a cycle in which students would resolve to avoid AI, only to return to it again under the time pressures of coursework and deadlines. For example, P19 described this pattern:

\begin{quote}
    \textit{``You know, the kind of \textbf{cycle of using it, feeling shame about it}, then saying, `I'm not going to use it anymore', but then using it again because you just have to turn this assignment in. \textbf{You slowly build up a version of your life that can't exist without it}."} (P19)
\end{quote}

\noindent Not wanting to use AI, but feeling powerless to avoid using it, was a common sentiment shared by participants. P19 shared how this was part of a broader problem related to self-regulation, saying, 

\begin{quote}
    \textit{``There's multiple vectors on which that \textbf{impulsivity is being bred into people now}...I'm addicted to my phone too, like I think it's not just AI.''} (P19)
\end{quote}

\noindent Similarly, P17 offered some justification for these shame spirals. For P17, it was driven by a feeling of endless output and pressure to keep up, using the metaphor of ``drowning'' to convey his overwhelm:  

\begin{quote}
    \textit{``\textbf{I feel this shame and guilt for maybe a couple of days}. After that, \textbf{I have to move on} to the next assignment and all the other schoolwork, and I'm just drowning in more work and school stuff. So it usually goes away. \textbf{But then it comes back if I use it again.}''} 
\end{quote}

\noindent Many shared this sentiment of feeling like AI was their only option to keep up with the work. P5 described trying to solve it first, but then relying on AI when they got stuck, 

\begin{quote}
    \textit{``It's like \textbf{a never-ending cycle}. I know that I shouldn't do this, I should try getting it done by myself first ... But then again, \textbf{I'll be doing the same thing over and over again}, this is like a really bad cycle.''} (P5)
\end{quote}

\noindent The lack of perceived agency in ending these cycles was a common thread in the interviews. Multiple participants believed they did not have a choice and felt powerless to stop their AI use:

\begin{quote}
    \textit{``I do find myself \textbf{stuck in that spiral} where I need to use it. I have no other choice but to use this.''} (P15)
\end{quote}

\noindent It appeared that for some students, the more they used these tools, the less agency they perceived they had in completing the work on their own. This reinforced the pattern where they would then again rely on AI to get through the follow up work. In this way, shame did not function as a deterrent to use, but instead became part of a self-reinforcing cycle of reliance, concealment, and self-critique.

\subsubsection{Framing AI with Metaphors of Addiction}

To make sense of these cycles, several participants spontaneously described their experiences using the language of addiction. The word `addiction' was never introduced by interviewers. Students also did not describe addiction as enjoyment or engagement, but rather as a loss of control over their own behavior. Participants often expressed the idea that their use harmed their learning, but they felt powerless to regulate their use. For example, P15 describes this tension between their intentions and actual behaviors,

\begin{quote}
    \textit{``\textbf{It's like an addiction}...When you use AI you get that response back immediately, you find yourself in that cycle knowing that it's wrong, you shouldn't do it, it's not going to help you in the long run.'' } (P15)
\end{quote}

\noindent Similarly, P19 compared AI use to more traditionally recognized forms of addiction, while adding that we don't yet have ways of talking about these experiences, 
\begin{quote}
    \textit{``We talk about addiction specifically with regard to things like \textbf{drugs and alcohol and sex}... we almost don't know how to talk about being addicted to AI.''} 
\end{quote}

\noindent A few participants consistently described AI as having it's own agency that actively undermined their ability to self-regulate their learning. Students used common metaphors in addition, such as a \textit{`devil on your shoulder,'} which P6 and P8 both used to describe this experience, for example 

\begin{quote}
    \textit{``I have this fear that it'll always just be like the \textbf{evil devil on your shoulder} saying \textbf{`Use me, use me, I can help.'}''} (P6)
\end{quote}
\begin{quote}
   \textit{``And so it's like...a \textbf{little like devil on your shoulder} saying, 'why don't you just use Gen AI and just get the grade that you want and you don't even have to put in the time and then you can go do other things.'' (P8)}
\end{quote}

\noindent This metaphor externalizes and personifies AI as an intrusive voice encouraging short-term relief that went against their longer-term learning goals and undermined their self-regulation.

\subsection{Fear and Apathy}

When asked about their potential future AI use, students frequently expressed fear, particularly in relation to its impact on their career paths. For instance, P7 said, 
\begin{quote}
    \textit{``Some people argue about how \textbf{AI is going to like take all of our jobs}, etc. I think a lot of people our age are \textbf{very pessimistic about the future in this way.}''} 
\end{quote} 

\noindent They went on to say that they notice people use this pessimism as a way to justify their AI use:  
\begin{quote}
    \textit{``I think it's just to say 'Oh, \textbf{I'm never going to get a job} ... so I'm going to give up.' That attitude ... \textbf{leads to more AI usage }since it's an\textbf{ easy way to get out of actually having to do work.}''} (P7)
\end{quote} 

\noindent Additionally, fear about the future and AI use exemplified how some students are developing a focus on outcomes over process: 

\begin{quote}
    \textit{``I think kids are just getting a bit lazy. They're using AI because they're like, \textbf{`You know what? I don't care. I'll be able to use AI in the future anyways.} It doesn't really matter. I just want to get the degree.'''} (P9)
\end{quote}

\noindent P19 says they get the sense that young people are nihilistic because of the state of the world and their futures.
\begin{quote}
    \textit{``A lot of older people talk about Gen Z as...  apathetic and indifferent... there is [also] that kind of engagement with AI.''}
\end{quote}

\noindent P14 even said he felt so worried about a future career, he was thinking of switching majors. 
\begin{quote}
    \textit{``It makes me think, \textbf{do I actually want to major in computer science?} It's my interest... but I don't know. [AI] makes me question a lot. Like, if it'll work out, or if I should switch to something different, like IS\&T, cybersecurity.''} (P14)
\end{quote}

Participants often expressed pessimism about their learning and future careers. These sentiments often appeared alongside beliefs that their personal efforts may not meaningfully impact their learning or career outcomes. In expectancy-value terms~\cite{wigfield2000expectancy}, these instances suggest a shift in how students consider investing effort in learning computer science. Historically, computer science required significant effort, but for many the effort was worthwhile. This may help explain why some students disengage from the learning process while others simply intend to pursue credentials.

\section{Discussion}

Our findings suggest that feelings of shame and guilt are not simply emotions that students experience around their AI use, but actually shape how students relate to AI tools, to others, and to themselves. 

\begin{quote}
    \textit{``Just talking about [AI use] now is \textbf{making me have more realizations} because I feel like it \textbf{never is talked about this deeply}.''} (P12)
\end{quote}

\noindent This students' reflection captures a core contribution of our work, that while AI use has become much more common~\cite{hou2025usage, prather2023robots, hou2024effects, alpizarChacon2025excited}, students emotional experiences with these tools, such as shame and guilt, remain under explored. Our work expands on social aspects of AI use~\cite{hou2025all, padiyath2024insights} by incorporating the socio-emotional aspects of AI use in computing education.

\subsection{Emotional Aspects of AI Use}

This work extends prior findings that suggest a broad acceptance of AI tools~\cite{hou2024effects, alpizarChacon2025excited, stone2025gen, hou2025usage, padiyath2024insights} by showing that outward acceptance may coexist with unspoken negative emotions. Students' emotional experiences with AI function as a mechanism that shapes how they engage with these tools, relate to others, and negotiate social norms. Emotions functionally serve to regulate behavior, but for many of our participants, these regulatory processes appeared to simultaneously produce unintended cycles of concealment, judgment, and diminished agency. Across our data, students described over-relying on AI despite discomfort, often accompanied by a perceived loss of control. These dynamics suggest that emotional responses to AI use do not reliably support self-regulated use. At the same time, these experiences are embedded within social settings where AI use is both ubiquitous and stigmatized. Students described judging themselves, each other, and in one case, a faculty member for using AI. Students selectively disclose or conceal their use, shaping how norms are perceived and discussed.

\subsubsection{Shame, Concealment, and Social Regulation}

Our findings suggest that shame and guilt function less as regulators of AI use itself and more as regulators of the \textit{visibility} of AI use. Students described selectively disclosing or hiding their use depending on context, audience, and anticipated judgment. This included hiding their use from peers and instructors, avoiding explicit discussion of AI use, and actively managing how their behavior might be interpreted. Participants reported engaging in concrete concealment practices (e.g. switching tabs, dimming screens, or avoiding situations where their use might be observed). These behaviors align with prior work on shame and its association with withdrawal, concealment, and persistent negative-self evaluation, which can entrench negative cycles instead of leading to behavior change~\cite{tangney2003shame}.

While the functionalist view of guilt and shame describe how these emotions guide behavior by signaling norm violations and motivating corrective actions~\cite{barrett1995function, hooge2011function}, participants in our study often continued using AI tools despite discomfort, guilt, or self-criticism. Rather than resolving these tensions, shame and guilt appeared to primarily influence how students navigated the visibility of their use. Because AI use was often concealed or selectively disclosed, students do not have a balanced or complete view of how others are using AI tools. As a result, students often make assumptions about the frequency and legitimacy of how their peers use AI tools. This pattern is consistent with pluralistic ignorance~\cite{miller1987pluralistic} though we did not directly assess norm perception. When use is concealed, students may misperceive peer norms, leading them to over- or under-estimate AI use. 

Prior work on social shaping in computing education suggests that students' AI use is influenced by observing how others use AI~\cite{padiyath2024insights}. However, our findings suggest that what is observable is not simply a reflection of behavior, but it is actively curated through emotional regulation and impression management.

\subsubsection{Stigma, Identity, and Norm Perception} 

In computing education, students' experiences are deeply tied to their computing identity~\cite{grosse2023identity}. This includes their perceptions about belonging~\cite{tonso2006student}, competence~\cite{mahadeo2020developing, falkner2015gender}, and legitimacy (i.e.: recognition) as computing practitioners~\cite{hazari2010connecting, mahadeo2020developing, cheryan2009ambient, falkner2015gender}. These perceptions are strongly influenced by socially constructed norms around what it means to be a `good' computing professional~\cite{cheryan2009ambient}.

Our findings extend this perspective by showing that AI use has become entangled with identity and moral evaluations of competence, effort, and legitimacy as reflected by how students often framed their experiences through stigmatized identities. Participants worried about being perceived as one of ``those students'' who excessively rely on AI tools. Students described themselves and others as `frauds', `cheaters', and not ``real programmers'' for relying on AI tools. Several students described feeling guilty that they had not worked hard enough, while others questioned whether relying on AI undermined their legitimacy as computing students altogether. 

These moral evaluations were not only internal. Participants described publicly judging peers, instructors, and broader patterns of AI use around them. In one case, a participant described being criticized by a family member for using AI tools. In another case, students collectively criticized a faculty member for using AI in the classroom. These instances suggest that AI-related stigma is also socially negotiated and reinforced through peer judgment. 

These findings also help contextualize the concealment practices that we observed. By hiding their AI use, students can actively manage the risk of being interpreted through these stigmatized identities. Concerns about how AI use might be perceived influenced students' willingness to seek help, participate openly, or discuss their learning practices with instructors and peers. For some participants, this resulted in withdrawal from opportunities such as not attending office hours for fear of being judged. 

In summary, our findings suggest that AI use in computing education has become intertwined with students' identities as competent, hard-working, and legitimate computer scientists. In addition to being socially regulated through norms and visibility, the identities students fear being associated with also shape their behaviors.

\subsubsection{Cycles of AI Use and Self-Regulation}

Self-regulated learning is a cyclical process of setting goals, monitoring progress, and adjusting strategies as needed~\cite{zimmerman2002becoming}. Effective self-regulation depends both on recognizing discrepancies between intended and actual learning behaviors, and the ability to translate that awareness into positive behavioral change.

Prior work by \citet{prather2024widening} has shown that when students use AI they may experience an ``illusion of competence,'' where they overestimate their ability to use these tools effectively. This suggests one way that AI systems can disrupt learners' ability to accurately interpret and respond to their own learning progress.

Our findings extend this line of work by showing that self-regulatory breakdowns are not limited to a misalignment in students' perceptions of their AI use. Participants frequently described difficulty maintaining changes in their AI use over time, even when they recognized tension between their intended and actual AI use. Instead of stabilizing, participants described recurring cycles of AI use, reflection, attempted restraint, and subsequent return to use. Several participants described these experiences using metaphors of addiction, highlighting how they made sense of repeated engagement with AI tools despite intentions to reduce use.

These patterns suggest that self-regulatory processes remain active but often co-occurred with shame and guilt, which contributed to increased self-monitoring without positive behavioral change.

\subsection{Implications for Pedagogy}

Based on our findings, we offer the following implications:

\subsubsection{Avoiding Moralizing AI Use}

Students appear to be navigating AI use within moralized and stigmatized frames. Many of our participants internalized AI use as a marker of personal inadequacy, describing feelings of being \textit{`lazy', `dumb'}, or \textit{`dishonest.'} These self-directed judgments were reinforced through peer discussions that often stigmatized AI use. In one case, students shamed their instructor for using AI. Together, these dynamics contributed to a classroom environment in which AI use was stigmatized despite explicit permission by instructors. In contexts with highly restrictive or detection-focused AI policies~\cite{hoq2024detecting, liu2025differences}, this perception could be amplified. By centering enforcement and surveillance, such policies risk framing AI use as a violation to be caught, rather than a practice to be understood and developed. This framing may also feel inauthentic given industry expectations to use AI~\cite{mejia2025bridging}. 

Educators cannot assume that moral discomfort will naturally promote healthy engagement with AI. Because shame is cyclical and often reinforces secrecy rather than productive behavior~\cite{slepian2020shame}, instructors should be cautious about approaches that implicitly frame AI use as evidence of laziness, incompetence, or moral failure.

\subsubsection{Support open dialogue and co-design shared norms}

Our results suggest that students often experience AI use as a threat to their identity as a `good' student, which prompts selective disclosure and concealment. This reluctance to openly discuss AI use reinforces uncertainty around social norms, making it harder for instructors and peers to negotiate shared expectations. It can also lead to anticipatory compliance, where students follow their own interpretation of rules and expectations that have never been explicitly established. While the ongoing ambiguity and inconsistency around AI policies and norms~\cite{stone2025gen, prather2023robots} may be exacerbating this anticipatory compliance, students expressed concealment, shame, and threats to identity even in classroom settings where they described AI policies as being highly permissive. 

Ultimately, a better path forward may be for instructors to explicitly co-design AI policies with students in their classes. Through open-dialogue about responsible AI use, preferred AI policies can be collectively negotiated. Making policies and expectations explicit may also reduce misperceptions about AI use, limit shame-driven concealment, and support use that aligns with the learning goals.

\subsubsection{Supporting Students in Navigating Pressures and Moral Dilemmas}

Our findings suggest that many students experience fear regarding their future career prospects. Students rationalized their AI use by describing various pressures that drive cycles of use and reinforce reliance. Time pressure has historically been a primary driver of academic dishonesty~\cite{sheard2011computing, kann2025students, albluwi2019plagiarism} which has also been observed in AI contexts, such as when students choose to disable optional guardrails leading up to a deadline~\cite{kapoor2026exploring}. Students now additionally grapple with an unclear return on investment of their time. Students wrestled with questions about whether to continue majoring in computer science, or whether their struggle was justified if they will just rely on AI in the future. If not interrupted, this line of thinking may lead to a moral disengagement~\cite{bandura2011moral} where students rationalize that \textit{`everyone does it', `everything is broken'}, and \textit{`it won't matter in the future.' }

By acknowledging these realities, instructors can foster more honest reflection and support students in making intentional choices about AI use rather than reacting to external pressures. Additionally, supporting students in these environments may require reinforcing opportunities for persistence and mastery, helping students regain both agency and optimism toward learning~\cite{bandura1997self}. Reducing time pressures through scaffolding, flexible deadlines~\cite{kim2024student}, or structured support may also help restore students' sense of agency, as well as support more intentional use of AI rather than reactive reliance.

\section{Limitations}

This study is a qualitative, exploratory investigation based on semi-structured interviews. The purpose of this inductive approach was to surface the socio-emotional aspects AI use rather than testing hypotheses or making definitive, generalizable claims. Relatedly, this sample of students in North America is not intended to be representative of all students, cultures, or institutional contexts.

This study relies on self-reported experiences with AI. Given the sensitive nature of discussing shame and guilt, participants may have underreported, reframed, or unevenly disclosed aspects of their behavior despite assurances of confidentiality. To mitigate potential power dynamics (e.g. being interviewed by faculty), interviews were conducted by trained peer interviewers who explicitly avoided imposing normative judgments about AI use. As such, accounts reflect participants' interpretations of their emotions, and are reported as illustrative and preliminary, with an emphasis on understanding participants' lived experiences. We intentionally did not collect academic performance data, as requesting such information could have heightened evaluative concerns and reduced participants' comfort sharing candid reflections. However, this limits our ability to examine how experiences of shame, guilt, or reliance on AI may vary across students with different academic standing. 

Although a functionalist lens helped guide our analysis, it may not fully capture the identity aspects uncovered in our findings. Emotions such as shame and guilt were intertwined with judgments of being `not real programmers,' which extends beyond emotional aspects. Future work could explore how AI-mediated learning impacts students' sense of identity and status within computing. 

Finally, these findings should be interpreted within a rapidly evolving socio-technical context where the norms around AI use in education continue to emerge and evolve~\cite{hou2024effects, hou2025all, padiyath2024insights, alpizarChacon2025excited, zastudil2023generative, lau2023from}.

\section{Conclusion} 
We conducted 19 semi-structured interviews to explore computing students' relationships with their AI use through a functionalist lens of shame and guilt. Our findings suggest that shame and guilt, and the behaviors that follow, have important implications for students' academic identity, confidence, and self-perception. Students described experiencing shame, engaging in hiding behaviors, and in some cases, perceiving their AI use as addictive. These experiences emerged alongside broader feelings of nihilism and pessimism. Our findings suggest that prevailing responses to AI use, particularly those emphasizing surveillance and detection, risk overlooking or exacerbating socio-emotional costs of AI use. Left unaddressed, these dynamics may undermine students' learning, identities, and long-term engagement in computing fields.

\balance

\newpage
\bibliographystyle{ACM-Reference-Format}
\bibliography{sample-base}

\end{document}